\documentclass[12pt]{article}

\usepackage{amssymb,latexsym,amsfonts,amsmath,url}
\def\Exp{{\mathbb{E}}}

\newcommand\independent{\protect\mathpalette{\protect\independenT}{\perp}}
\def\independenT#1#2{\mathrel{\rlap{$#1#2$}\mkern2mu{#1#2}}}

\newcommand{\Z}{{\mathbb Z}}
\newcommand{\R}{{\mathbb R}}
\newcommand{\N}{{\mathbb N}}

\newcommand{\cS}{{\cal S}}

\newtheorem{Definition}{Definition}
\newtheorem{Lemma}{Lemma}
\newtheorem{Theorem}{Theorem}

\title{On the entropy production of time series with unidirectional linearity}

\date{August 13, 2009}

\author{Dominik
Janzing\thanks{e-mail:
dominik.janzing@tuebingen.mpg.de}
\\ \small Max Planck Institute for Biological Cybernetics\\
\small  Spemannstr.~38\\
\small  72076 T\"ubingen, Germany 
}

\begin{document}
\maketitle

\begin{abstract}
There are non-Gaussian time series that admit a causal linear autoregressive moving average   (ARMA) model
when regressing  the  future on the past, but not when regressing the past on the future.
The reason is that, in the latter case, the regression residuals are only {\it uncorrelated}
but not statistically independent of the future. 
In previous work,  we have experimentally verified that many empirical time series indeed show such a time inversion 
asymmetry. 

For various physical systems,  it is known that time-inversion asymmetries are linked to the
thermodynamic entropy production in non-equilibrium states. Here we show that 
such a  link also exists for the above unidirectional linearity.

We study the  dynamical evolution of a physical toy system with linear coupling to an infinite environment 
and show that  the linearity of the dynamics is inherited to the  forward-time conditional probabilities, 
but not to the backward-time conditionals.
The reason for this asymmetry between past and future is that 
the environment
permanently provides particles that 
are in a product state before they interact with the system, but show statistical dependencies
afterwards.
From a coarse-grained perspective,
the interaction thus  generates entropy. 
We quantitatively relate the strength of the non-linearity of the backward conditionals
to the minimal amount of entropy generation.    
\end{abstract}

\section{Unidirectional linearity in time series}

To study the implications and the different versions of the thermodynamic arrow of time has attracted interest of theoretical physicists and philosophers since a long  time \cite{Reichenbach,penrose62direction,Balian,Gibbs,JaynesGibbs,Lebowitz,Wallace}. 
More specificly, it is the question how the difference between time reversibility of microscopic 
physical dynamics is consistent with the existence of irreversible processes on the macroscopic level.
%It is commonly agreed that irreversibility of processes 
The most prominent examples of irreversibilities
(e.g. heat always flows from the hot to the cold reservoir, never vice versa,
every kind of energy can be converted into heat, but not vice versa)
can directly be explained by the fact that the processes generate entropy 
and their inverted counterpart is therefore forbidden by the second law. 

Here we describe an asymmetry between past and future whose connection to the second law is more subtle.  
An extensive analysis of  more than 1000 time series \cite{ICMLTimeSeries} showed that there are many cases
where
the statistics could be better explained by a linear autoregressive model 
from the past to the future and only few cases where regressing the past on the future yields a better model \cite{Jonas1,ICMLTimeSeries}. In the context of non-equilibrium thermodynamics it has been shown for various physical
models (e.g. \cite{MaesEx,Gallavotti}, and also in a  more abstract setting \cite{Maes}) that statistical asymmetries between past and  future can be related 
to thermodynamic entropy production. 

This paper is in the same  spirit,
but  we  will  try to use only those assumptions about the underlying physical system that are necessary
to make the case and try to simplify the argument as much as possible. 
The ingredients are 
(1) a system interacting  with an environment consisting of infinitely many subsystems that are initially in a product state, each system having 
an abstract vector space as phase space, (2)
linear volume preserving dynamical equations for the joint system. 
We will not refer to any other ingredients from   physics, 
like energy levels, thermal Gibbs  states, etc. Of course, this  raises the  question of how to define
 entropy production. Here, we interpret the generation of dependencies among an increasing number  of particles
this way.

To describe the model more precisely, 
we start with preliminary remarks on statistical dependencies.
First we introduce the following terminology.

\begin{Definition}[linear models]${}$\\
The joint distribution $P_{X,Y}$ of two real-valued random variables $X$ and $Y$
is said to admit a linear model $X\rightarrow  Y$ with additive noise (linear model, for  short) if  
$Y$  can be  written as
\[
Y:=\alpha X +\epsilon
\]
with a structure coefficient $\alpha \in \R$ and a noise  term  $\epsilon$ that is statistically independent of $X$ ($X \independent \epsilon$, for short).
%Since this is actually a property of $P(Y|x)$  we also say that $P(Y|X)$ admits a linear model.
\end{Definition}

It should be emphasized that statistical independence between two random variables $Z,W$ is defined by
factorizing probabilities
\[
P_{Z,W}=P_Z \otimes  P_W\,,
\]
instead of  the weaker  condition of uncorrelatedness, which is defined by  factorizing expectations:
\begin{equation}\label{uncorr}
\Exp(ZW)=\Exp(Z)\Exp(W)\,.
\end{equation}
Uncorrelatedness between $X$ and $\epsilon$ is automatically satisfied if $\alpha$  is chosen to minimize
the least square error. 

Except for the trivial case of independence, $P_{X,Y}$ can only admit linear models in both directions at  the same time if it is bivariate Gaussian. 
This can be shown using the theorem of Darmois Skitovich \cite{Darmois},  which we rephrase
now because it will also be used later.

\begin{Lemma}[Theorem of Darmois \& Skitovich]${}$\\
\label{Darmois}
Let $Y_1,Y_2,\dots,Y_k$  be statistically independent random variables and the two linear  combinations
\begin{eqnarray*}
W_1&:=&\sum_{j=1}^k \beta^{(1)}_j Y_j\\
W_2&:=&\sum_{j=1}^k \beta^{(2)}_j Y_j
\end{eqnarray*}
be independent. Then all $Y_j$ with $\beta^{(1)}_j \beta_j^{(2)}\neq 0$ are Gaussian.
\end{Lemma}

% in  \cite{Hoyer,ICMLnonlingam} 
%we  studied a generalization where the effect is a function of
%the cause up to some  additive noise that is independent of the cause. 
%Due to the sucess of such causal inference methods, it would be desirable
%to further justify them by understanding the relation to the second law.

In  the context  of
causal inference from statistical data, it has been proposed to consider the  direction of the linear  model
as  the {\it causal} direction \cite{Kano2003,Shimizu2005}.
In \cite{ICMLTimeSeries} we have shown that the same idea can be used to solve the following  binary classification  problem: Given numbers $X_1,X_2,X_3,\dots$ 
that are  known to be the values of an empirical time series in their correct or in their time reversed  order. 
Decide whether $X_1,X_2,X_3,\dots$ or $\dots,X_3,X_2,X_1$ is the correct order. 
Certainly, this problem is less relevant  
than the problem of inferring causality since our experiment required to artificially blur the true direction  even though it was actually known.
The motivation for our study was to test causal inference principles
by applying them to this artificial problem.

To explain our ``time direction inference  rule'' we 
first introduce an important class of stochastic processes \cite{Brockwell}:

\begin{Definition}[ARMA models]${}$\\
We call a time series $(X_t)_{t\in \Z}$  an autoregressive moving average process 
of order $(p,q)$ if it is weakly stationary and there is an iid noise $\epsilon_t$ with mean zero such that
\[
X_t=\sum_{i=1}^p \phi_i X_{t-i} +\sum_{j=1}^q \theta_j \epsilon_{t-j} +\epsilon_t  \quad  \forall t\in \Z\,.
\]  
For $q=0$ the process reduces to an autoregressive process and for $p=0$ to a moving  average process.
The short-hand notations are $ARMA(p,q)$, $AR(p)$, and $MA(q)$. 
The first  and the second sums are  called the AR-part and the MA-part, respectively. 

The process is called causal\footnote{\cite{Brockwell} chooses a different definition, but 
we have  argued in \cite{ICMLTimeSeries} that
it is equivalent to ours.}
if
\begin{equation}\label{IndCon}
\epsilon_t \independent X_{t-i}   \quad \forall i>0\,.
\end{equation}
\end{Definition}

\noindent
Note that  a process is  called weakly stationary if the mean $\Exp(X_t)$ and second order moments
$\Exp(X_t X_{t+h})$ are constant in time \cite{Brockwell}.
In \cite{ICMLTimeSeries} we  have shown the  following theorem: 

\begin{Theorem}[non-invertibility of non-Gaussian processes]${}$\\
\label{Jonas}
If $(X_t)_{t\in \Z}$ is a causal ARMA process with non-vanishing $AR$-part, then
$(X_{-t})_{t\in  \Z}$  is a causal ARMA process if and only if $(X_t)$ is a Gaussian process.
\end{Theorem}

In particular, a process with  long-tailed distributions like e.g.~Cauchy can only be causal in one direction
(provided that it has an AR-part). 
In \cite{ICMLTimeSeries} 
we have postulated that whenever a time series has a causal ARMA model in one direction but not the other
the former is likely to be the true one, but some remarks on the practical implementation need to be made:
Testing condition (\ref{IndCon}) yields p-values for the hypothesis of independence.
The performance of our inference method depends heavily on how these p-values are used to
decide whether a linear model is accepted for one and only one of the directions. Our rule depends on  two parameters
$\alpha$ and $\delta$, the significance level and  the gap, respectively.
We say that an ARMA model is accepted for one direction but not the other if the 
p-value for the  direction under consideration is above $\alpha$ and it  is below $\alpha$ for  the converse direction and, moreover, the
gap is at least $\delta$.
By choosing a small value $\alpha$ and a large value $\delta$ one gets fewer decisions but  also the  fraction
of wrong classifications decreases. 
On
1180 empirical time series from EEGs  \cite{ICMLTimeSeries} we where able to classify around $82 \%$ correctly when the parameters are set to yields decisions for about $4\%$ of the time series. When decisions were made for 
a larger fraction of time series, the number of correct answers still significantly exceeded chance level. 
Qualitively similar results were obtained for 200 time series from different areas, like finance, physics, transportation, crime, production of goods, demography, economy, Neuroscience, and agriculture \cite{Jonas1}.

\section{Physical toy model}

%The following model gives an idea of how the observed asymmetry is linked to the production of coarse-grained entropy 
% in the environment.

Here we describe a physical model that suggests that the observed asymmetry is an implication of generally accepted asymmetries between past and future.
We assume that the values $X_t$ as observables of a classical physical system.\footnote{Of course, such an embedding
is hard to imagine for time series from stock markets, for instance. However, other time series, e.g., 
EEG-data, are closer related to {\it physical} observables.}
For our toy model, we use only two properties of physical models that we consider 
decisive for the argument:\\
(1) The state of a system is a point in some phase-space $\cS$ that is a sub-manifold of $\R^n$. \\
(2) The dynamical evolution of an isolated system is given by a family $M_t$ of 
volume-preserving 
bijections on $\cS$. \\
Due to Liouville's Theorem, this holds for the 
dynamics of all Hamiltonian systems, 
other dynamical maps can only be obtained by restricting the joint evolution of
a composed system to one of its components.

For simplicity, we restrict the attention to an $AR(1)$ process:
\begin{equation}\label{AR}
X_t=\phi X_{t-1} +\epsilon_t\,. 
\end{equation}
We will now interpret $X_t$ as a physical observable of a system
$S^{(0)}$, whose state is changed by interacting with its environment.
The latter consists of an infinite collection of subsystems $S^{(j)}$ with  $j\in \Z\setminus \{0\}$.
%Later, $E^{(-1)},E^{(-2)},\dots$ and $E^{(1)},E^{(2)},\dots$ will be interpreted as incoming and outgoing particles,
%respectively. 
Each subsystem is described by the
real-valued observable $Z^{(j)}$. 
Its value at time  $t$ is denoted by $Z^{(j)}_t$, hence $X_t=Z^{(0)}_t$, but we
will keep the notation $X_t$ whenever its special status among the variables should be emphasized.

Then we define a joint time evolution  by
\begin{eqnarray}
Z^{(0)}_{t+1}&=&\gamma_{11} Z^{(0)}_t +\gamma_{12} Z^{(-1)}_t \label{dyn1}\\ 
Z^{(1)}_{t+1} &=& \gamma_{21} Z^{(0)}_t +\gamma_{22} Z^{(-1)}_t \label{dyn2}\\
Z^{(j)}_{t+1} &=& Z^{(j-1)}_t  \quad \hbox{ for } j\neq 0,1 \label{dyn3}\,.
\end{eqnarray}
The dynamics thus is a concatenation of 
the map $\Gamma$ on $\R^2$, given by the entries $\gamma_{kl}$,
with a shift propagating the state of subsystem $S^{(j)}$ to  $S^{(j+1)}$.

The environment may be thought of as a beam of particles that approaches site $S^{(0)}$, interacts with it,
and disappears to infinity; we  have discretized the propagation only to make it  compatible with the
discrete stochastic process. The interaction is given by $\Gamma$.
The phase space of  the systems $S^{(j)}$ may be larger than one-dimensional, but we assume that the variables
$Z^{(j)}_t$ define the  observables that are relevant for the interaction.
To ensure conservation of volume in the entire phase space,
$\Gamma$ needs to be volume-preserving, i.e. $|{\tt det}(\Gamma)|=1$.
Since our model should be interpreted as the discretization of a continuous time process we assume
$\Gamma\in SL(2)$.  

One checks easily that the above dynamical  system generates for $t>0$ the causal $AR(1)$-process
\[
X_{t}=\gamma_{11}X_{t-1} +\epsilon_t \quad \hbox{ with } \quad \epsilon_t:=\gamma_{12} Z^{(0)}_t\,,
\]
if we impose the initial conditions
\begin{equation}\label{InitialCond}
 Z_0^{(j)} \hbox{ i.i.d. with  some  distribution } Q
\end{equation}
Actually, it would be sufficient to impose independence only for  the
non-positive $j$, but later it will be convenient to include also positive values $j$ and assume that the whole 
ARMA process has a starting time $t=0$. This will make  it easier to  track the increase of  dependencies over time.
The fact that every $Z_0^{(j)}$ is drawn  from the same distribution $Q$ ensures that the process $(X_t)_{t\in \N}$ is stationary.

We will now show that, under generic conditions, 
the dynamics creates statistical dependencies between the
subsystems. We will later see that this is the reason why
the time-inverted version of the above scenario 
would not be a reasonable physical
model
for the process $(X_{-t})$. We need the following Lemma:

\begin{Lemma}[dependencies from sequences of adjacent operations]${}$\\ \label{depCreat}
Let $\Gamma\in SL(2)$ have non-diagonal and diagonal entries. 
Denote by $\Gamma^{(n)}_{l,l+1}$ the embedding into the two-dimensional subspaces of $\R^n$  
that correspond to consecutive components $l,l+1$ with  $l=0,\dots,n-1$,  i.e., 
\[
\Gamma^{(n)}_{l,l+1}:={\bf  1}_{l-1} \oplus \Gamma \oplus {\bf 1}_{n-l-1}\,,
\]
where ${\bf 1}_m$ denotes the identity matrix in  $m$ dimensions.
Let $P$ be a non-Gaussian distribution on $\R$.
Then the application of
\[
\Gamma^{(n)}_{0,1}\circ \Gamma^{(n)}_{2,3} \circ \cdots \circ \Gamma^{(n)}_{n-2,n-1}
\]
to $\R^{n}$ turns the product  distribution  $P^{\otimes n}$ into a non-product distribution.
\end{Lemma}

\noindent
Proof: Due to Lemma~\ref{Darmois}, $\Gamma^{(n)}_{n-2,n-1}$ generates dependencies between the  last and the second last component. Since none of the other operations acts on the last component, the  dependence between
the last component and the joint system given by the remaining $n-2$ 
components, is preserved. $\Box$
\vspace{0.3cm}

To apply Lemma~\ref{depCreat} to our system, it is sufficient  to focus on the region of the
chain on  which  the dependencies have been generated  after the time $t$ under consideration. It  is given by
\begin{equation}\label{region}
S^{0,\dots,t}:=S^{(0)}\times S^{(1)}\times \cdots S^{(t)}\,.
\end{equation}
Its state is given  by the variable transformation
\begin{equation}\label{outgoing}
(Z^{(0)}_t,Z_t^{(1)},\cdots,Z^{(t)}_t)= (\Gamma^{(t+1)}_{0,1}\circ \Gamma^{(t+1)}_{1,2} \circ \cdots \circ \Gamma^{(t+1)}_{t-1,t}) 
(Z_0^{(-t)},\cdots,Z_0^{(0)})  
\,,
\end{equation}
and  all the other sites are still jointly independent and independent of the region  (\ref{region}).
If the relation between $X_t$ and  $X_{t+1}$ is non-trivial
 (i.e., neither deterministic nor independent)
$\Gamma$ must have diagonal and non-diagonal entries, which implies that
(\ref{outgoing}) is not a product state.

The following argument shows that
the dependencies between the outgoing  particles is closely linked to the irreversibility of the scenario:
The fact that the time evolution generates a {\it causal} $AR(1)$-process is ensured by independence of
$Z^{(0)}_t,Z^{(-1)}_t,Z_t^{(-2)},\dots$ describing the
incoming particles.
If the variables $Z^{(1)}_t,Z_t^{(2)},\dots$ are also independent we can run the process backwards
to induce the {\it causal} $AR(1)$-process $(X_{-t})$. However, by virtue of  Theorem~\ref{Jonas}, this is only possible for
$(X_t)$ Gaussian.

Summarizing the essential part of  the argument, the joint distribution $P_{X_t,X_{t+1}}$ has a linear model 
from $X_t$ to $X_{t+1}$ but not vice versa 
because the incoming particles are jointly independent 
but the outgoing  particles are dependent.
Now we show a quantitative relation between the non-linearity in backward time direction
and the generated dependencies.
To this end,
we measure the strength of  the statistical dependencies of the joint system as follows.
If a system consists of finitely many  subsystems its multi-information is defined by
\[
I(Y_1,\dots,Y_k):=\sum_{j=1}^k H(Y_j) -H(Y_1,\dots,Y_k)\,.
\]
Here,
$H(.)$ is the differential Shannon entropy
\cite{Cover}
\[
H(Y_1,\dots,Y_n):=-\int p(y_1,\dots,y_n)\log  p(y_1,\dots,y_n) dy_1, \cdots dy_n\,,
\]
where $p(y_1,\dots,y_n)$ denotes the joint probability density of the random variables
$Y_1,\dots,Y_n$. 
For $k=2$,  the  multi-information
coincides with the usual mutual information $I(Y_1:Y_2)$.

For our infinite system we define multi-information as follows:

\begin{Definition}[multi-information]${}$\\
The multi-information of the joint system of
all $S^{(j)}$ at time $t$ is defined by 
\[
I(t):= \lim_{m\to \infty} I_{-m,\dots,m}(t)\,,
\]
whenever the limit exists.
%Let $D$ be the dynamics given by eqs.~(\ref{dyn1}) to (\ref{dyn3})
%let the variables $X_{-1},X_{-2},\dots $ be jointly independent.
%Then the phenomenological entropy generated by $D$  at the transition $t\mapsto t+1$ is 
%given byDarmois $I( )$. 
\end{Definition}

%We will interpret the increase of multi-information as phenomenological increase of entropy, because
%it coincides with the increase of the sum of entropies over the subsystems.  
Its increase in time can easily be computed:

\begin{Lemma}[multi-information as pairwise information]${}$\\  \label{MIgen}
Let the initial state of $S^{-\infty \dots \infty}$ satisfy the conditions~(\ref{InitialCond}). 
Then the multi-information generated by the process
in eqs.~(\ref{dyn1}) to (\ref{dyn3}) with $\Gamma\in SL(2)$ satisfies:
\[
I(t)-I(t-1)= I(Z^{(0)}_t:Z^{(1)}_t)  \quad\forall t\geq 0\,.
\]
\end{Lemma}

\noindent
Proof:  
We consider the  state of the system $S^{0,\cdots,t}$ at time $t$ that we had obtained if the interaction
would have been inactive (i.e., $\Gamma={\bf 1}$) during  the last time step.
It  is described by
the transformed variables
\begin{equation}\label{modi}
(\tilde{Z}_{t}^{(0)},\cdots,\tilde{Z}^{(t)}_{t}):=
(\Gamma^{(t+1)}_{1,2} \circ \Gamma^{(t+1)}_{2,3} \circ \cdots \Gamma^{(t+1)}_{t-1,t}) (Z_0^{(-t)},\cdots,Z_0^{(0)})\,.
\end{equation}
Their  multi-information coincides with $I(t-1)$ because the shift part of the dynamics is irrelevant.

The true state of system $S^{0,\cdots,t}$ at  time $t$ is then given by additionally 
applying $\Gamma^{(t)}_{0,1}$ to eq.~(\ref{modi}).
The increase of  multi-information caused by applying  $\Gamma$ to system $S^{(0)}$ and  $S^{(1)}$ 
can be computed as follows.
Clearly,  the joint  entropy of the system   $S^{0,\dots,t}$ remains constant.  Hence  the 
only change of multi-information is due to the change of the marginal  entropies of 
$S^{(0)}$ and $S^{(1)}$. 
Since $\Gamma^{(t+1)}_{0,1}$ also preserves  the joint entropy of system $S^{0,1}$, 
the  increase of the marginal entropies coincides with the
pairwise mutual information created between $S^{(0)}$ and $S^{(1)}$. Hence, 
\begin{eqnarray*}
I(t)- I(t-1)&=&I(Z^{(0)}_{t}:Z^{(1)}_{t})
%&=&\sum_{j=0}^{t-1}  I(Z^{(j)}_t:Z^{(j+1)}_t) +
%I(Z^{(0)}_{t+1},Z^{(1)}_{t+1})\\
%&=& \sum_{j=1}^{t}  I(Z^{(j)}_{t+1}:Z^{(j+1)}_{t+1})+
%I(Z^{(0)}_{t+1},Z^{(1)}_{t+1})\\
%&=& \sum_{j=0}^{t}  I(Z^{(j)}_{t+1}:Z^{(j+1)}_{t+1})
 \,,
\end{eqnarray*}
where we have used  the fact  that the state of all systems
$S^{(j)}$ with $j>0$ is only shifted. 
$\Box$

%\begin{Theorem}[non-invertibility of the model]${}$\\
%Let $D$ be a dynamics according to eqs.~(\ref{dyn1}) to (\ref{dyn3}) that induces a  non-Gaussian 
%causal AR(1)-process. Then $D$ generates multi-information.
%\end{Theorem}

%\noindent
%Proof: 
%If the  process does not generate multi-information, $\Gamma$ must be chosen such that
%$X_{t}$ and $Z_t^{(1)}$ are  independent. Due to eqs.~(\ref{dyn1}) and (\ref{dyn2}), they are given by two
%different linear  combinations of $X_{t-1}$ and $Z_{t-1}^{(0)}$. If one of the variables
%is non-Gaussian, this can only be  the case if $\Gamma$  is diagonal or if $\gamma_{11}=\gamma_{22}=0$ (Lemma~\ref
%{Darmois}).
%The first case yields  a deterministic process, the second
%leads to
%independence between $X_t$ and  $X_{t-1}$.$\Box$ 

\vspace{0.3cm}
To show the link between the amount of generated dependencies and the non-linearity of the backward process,
we measure the latter as follows.

\begin{Definition}[measuring non-linearity of joint distributions]${}$\\
Let $L$ be the set of joint distributions $R_{X,Y}$ that admit a linear model from $X$ to $Y$.
Set
\[
D(P_{X,Y}||L) := \inf_{R_{X,Y} \in L} D(P_{X,Y}||R_{X,Y})\,,
\]
where $D$ denotes the relative entropy  distance \cite{Cover} 
and the infimum is taken  over all distributions in $L$.
% for which it is finite.
\end{Definition}

\noindent
Then we have:

\begin{Theorem}[non-linearity of backwards model and multi-inf.]${}$\\
Let $(X_t)$ be a causal $AR(1)$-process and $I(t)$ the multi-information
of all the ``particles''  in the toy model given by eqs.~(\ref{dyn1}) to (\ref{dyn3}).
Then,
\[
I(t)-I(t-1)\geq D(P_{X_{t},X_{t-1}}||L)\,.  
\]
\end{Theorem}

\noindent
Proof: 
Assume $X_t$ and $X_{t-1}$ are neither 
linear dependent nor statistically independent
because otherwise the bound becomes trivial  since we had
$
P_{X_{t},X_{t-1}}\in L
$.
The idea of  the proof is the following: we figure out how much the joint  distribution of
$X_t$ and $X_{t-1}$ has to be modified to admit a linear model from $X_{t}$ to $X_{t-1}$.
We have already argued that the entire stochastic process would admit a linear model 
in backward direction
if all the outgoing particles
were statistically independent. To obtain a linear model only from $X_{t}$ to $X_{t-1}$ 
by reversing the physical toy model
it is sufficient 
to replace $S^{(1)}$ at  time $t$ with a system that is independent of the remaining ones.
More precisely, we replace the joint distribution $P$ of all $Z_t^{(j)}$
 by the unique distribution $\tilde{P}$ for which $Z_t^{(1)}$ and  the remaining variables are independent
but the marginal distribution to $Z_t^{(1)}$ and the rest coincide with $P$, i.e.,
\[
\tilde{P}:=P_{Z_t^{(1)}}\otimes P_{ \dots,Z^{(-1)}_t,Z^{(0)}_t,Z_t^{(2)},Z_t^{(2)},\dots} \,.
\]
Then  
we check how this changes the joint distribution of $X_{t}$ and $X_{t-1}$.
The inverse dynamics $t\mapsto t-1$ is given  by  
\begin{eqnarray}
Z^{(0)}_{t-1}&=&\tilde{\gamma}_{11} Z^{(1)}_t +\tilde{\gamma}_{12} Z^{(0)}_t \label{dyn1rev}\\ 
Z^{(-1)}_{t-1} &=& \tilde{\gamma}_{21} Z^{(1)}_t +\tilde{\gamma}_{22} Z^{(0)}_t \label{dyn2rev}\\
Z^{(j)}_{t-1} &=& Z^{(j-1)}_{t}  \quad \hbox{ for } j\neq 0,-1 \label{dyn3rev}\,,
\end{eqnarray}
where $\tilde{\gamma}_{kl}$ denote the entries of $\Gamma^{-1}$.

Since $X_t=Z_t^{(0)}$ and 
\begin{equation}\label{trans}
X_{t-1}=\tilde{\gamma}_{11} Z_t^{(0)} +\tilde{\gamma}_{12} Z_t^{(1)} \,,
\end{equation}
which  is implied   by
 eq.~(\ref{dyn1rev}), 
the pair $(Z^{(0)}_t,Z^{(1)}_t)$ and $(X_t,X_{t-1})$ span the  same probability space
(note that both coefficients in eq.~(\ref{trans}) are  non-zero because we  have excluded
the  cases of linear dependency and statistical independence).
Hence 
$\tilde{P}_{Z^{(0)}_t,Z^{(1)}_t}$ induces by variable transformation a distribution 
$\tilde{P}_{X_t,X_{t-1}}$ satisfying
\[
D(P_{X_t,X_{t-1}}||\tilde{P}_{X_t,X_{t-1}})=D(P_{Z^{(0)}_t,Z^{(1)}_t}||\tilde{P}_{Z^{(0)}_t,Z^{(1)}_t})\,.
\]
The left hand  side is an upper bound  for the distance of $P_{X_{t-1},X_{t}}$  to a linear model
from  $X_t$ to  $X_{t-1}$ 
because $\tilde{P}_{X_{t},X_{t-1}}$ admits such a model. 
The right hand side coincides with the mutual information between
$Z^{(1)}_t$  and $X_t=Z^{(0)}_t$ (since mutual information is known to be the relative entropy distance to  the 
product of marginal distributions \cite{Cover}), which  is exactly the multi-information generated in step
$(t-1)\mapsto t$ due to Lemma~\ref{MIgen}. 
$\Box$

\vspace{0.3cm}
If $X_t$ is Gaussian, the stochastic process can be obtained without generation of multi-information: 
If $C$  denotes the covariance matrix of the pair $(X_t,Z^{(-1)}_t)$, which is diagonal by assumption (because 
the variables are independent and identically distributed),
then the generation of  multi-information is zero if and only if $\Gamma^T C \Gamma$ is diagonal.
The easiest case is that 
$\Gamma$ rotates the space  $\R^2$ by some angle $\alpha$. Even though this dynamics leaves the entire joint state 
of the system invariant, it
can induce any stationary 
AR(1)-process. This is because  then
$|\phi|^2\leq 1$ in eq.~(\ref{AR})  and
we can thus write
\[
X_{t+1}=\cos \alpha X_t +\epsilon_t
\]
with $\epsilon_t:= \sin \alpha Z^{(0)}_t$. 

Note that Gaussian processes can also be realized by a system that {\it does} generate multi-information.
For instance, 
\[
\Gamma:=\left(\begin{array}{cc} \cos \alpha & \sin \alpha \\ 0 & \cos^{-1} \alpha \end{array}\right)\,.
\]
induces the same process $(X_t)$ as a rotation by the angle $\alpha$, but induces
dependent outgoing  particles because $\Gamma^T \Gamma$ is non-diagonal. 
This shows that the correspondence between entropy production and  time-inversion 
asymmetry of $(X_t)$ can only consist of {\it lower} bounds.

\section{Interpretation}

We  first  discuss the interpretation of the Gaussian case.
To show an even closer link to thermodynamics, we recall that
Gaussian distributions often occur in the context of 
thermal equilibrium states.  For instance, the variable position and  momentum 
of a harmonic oscillator
are Gaussian distributed  in thermal equilibrium.
Hence we interpret the case of the isotropic Gaussian 
as thermal equilibrium dynamics.  The fact that the joint distribution
$P_{X_t,X_{t+1}}$ coincides with $P_{X_t,X_{t-1}}$ is exactly the symmetry imposed by the well-known
{\it detailed-balance} condition \cite{Tolman} that holds for every Gibbs  state.

In order to interpret the scenario in the non-Gaussian case as entropy production,
we note that the sum over the marginal  entropies of the subsystems increase linearly in time. The fact that 
the joint Shannon entropy remains constant loses more and more its practical relevance since it requires
complex joint operations  to undo the dependencies. 
From a coarse-grained point of view, the entropy increases in  every step. 

%To further discuss why the time  inverse of our scenario is unphysical, we

%We have thus shown that our linear joint dynamics induces processes that admit a linear  model 
%from $X_t$ to $X_{t+1}$ but not from $X_{t+1}$ to $X_t$ and relates the strength of the non-linearity to
%the thermodynamic irreversibility of the process.
%In the next section we will further discuss the difficulty of undoing the dependences again

In our experiments we found several examples of time series 
that could better be fit with a causal ARMA model from the future to the past
than vice versa, even though this was only a minority of those for which a decision was made.
Of course, there is no contradiction to the second law if this is the case.
To avoid such misconclusions we discuss which assumptions could be violated 
to generate time series that admit non-Gaussian ARMA models in the {\it wrong} direction.

To this end, we list  the requirements which jointly  
make the time-inverted scenario of the above dynamics extremely unlikely:

\begin{enumerate}

\item The ``incoming particles'' (which correspond to the outgoing ones in the  
original scenario)
and $S^{(0)}$ had  to be statistically dependent.\footnote{This indicates that they have already been interacting earlier, cf. Reichenbach's principle of the common cause \cite{Reichenbach}, which
is meanwhile one of the cornerstones of
causal inference} 

\item 
The coupling between $S^{(0)}$ and the incoming particles must be chosen such that it exactly removes the
incoming dependencies. There is nothing wrong with {\it dependent} particles approaching $S^{(0)}$, and a coupling that
destroys dependencies between the particles and $S^{(0)}$ by creating additional dependencies with a third party.
However, removing dependencies in a {\it closed} system requires transformations that  are specificly adapted
to the kind of dependencies that  are present.
In other  words, the coupling between $S^{(0)}$ and the incoming particles
had  to be one of the ``few'' linear maps $\tilde{\Gamma}\in SL(2)$ needed for undoing the operation that created the
incoming dependencies. 

\end{enumerate}

We want to be  more explicit about the last item  and 
recall that the joint state of $S^{0,\cdots,t}$ at  time $t$ is  given  by 
\[
(\Gamma^{(t+1)}_{0,1}\circ \Gamma^{(t+1)}_{1,2}\circ \cdots \circ \Gamma^{(t+1)}_{t-1,t}) Q^{\otimes (t+1)}\,.
\]
We  now run  the time inverted dynamics (\ref{dyn1rev})--(\ref{dyn2rev})
(starting from $t$ and ending at $0$)  
to this input 
using some 
arbitrary $\tilde{\Gamma}\in SL(2)$. 
The state of $S^{-t,\cdots,0}$  then  reads
\[
(\hat{\Gamma}^{(t+1)}_{0,1} \circ \hat{\Gamma}^{(t+1)}_{1,2}\circ \cdots \circ \hat{\Gamma}^{(t+1)}_{t-1,t})  Q^{\otimes (t+1)} \,,
\]
where  we  have defined
\[
\hat{\Gamma}:=\tilde{\Gamma} \circ \Gamma\,.
\]
Due to Lemma~\ref{depCreat}, this can only be a product state if $\hat{\Gamma}$ has only diagonal or only off-diagonal entries (or if $Q$ is   Gaussian).
This shows that the dependencies can only be resolved by $\tilde{\Gamma}$ if it is adjusted  to
the specific form of the dependencies of the incoming particles.

This kind of mutual adjustment between mechanism
and incoming state is unlikely. Similar arguments have been used in causal inference recently \cite{LemeireD,Algorithmic}. According to the language used there, the incoming state and the coupling
share {\it algorithmic} information, which indicates that the incoming state and the coupling have not been chosen independently.\footnote{Note that the thermodynamic relevance of {\it algorithmic} information
has also been pointed out in \cite{ZurekKol}.}

To generate a process $(X_t)_{t\in \Z}$ that admits a linear model in backward direction thus requires 
a different class of dynamical models. For instance, the joint dynamics could be non-linear.

\section{Conclusions and discussion}

We have discussed time series  that admit a causal ARMA model in forward direction but requires non-linear transitions
in backward directions to remain causal. Since previous experiments verified that some empirical time series
indeed show this asymmetry, we have presented a model  that relates it to the thermodynamic arrow of time.

To this  end,
we have presented a toy model of a physical system coupled to an infinite environment where we linked the
asymmetry to the thermodynamical entropy production.

The essential point is that the linearity of the joint dynamics is inherited to the forward
but not to the backward conditionals. Of course, not every physical dynamics is linear.
Nevertheless, the result suggests that simplicity of the laws of nature 
is  inherited only to the forward time conditionals. Since stochastic processes
usually describes the state of a system that strongly interacts with its environment
there is no simple entropy  criterion to distinguish between the true  and  the wrong time direction. Hence, more 
subtle asymmetries as the  ones described here are required.

The asymmetries fit to observations  in \cite{OccamsRazor}
discussing physical interacting models of a causal relation between two random variables
$X$  (cause) and $Y$ (effect), where $P(Y|X)$ was simple and $P(X|Y)$  complex,  which has  been used in recent
causal inference methods \cite{Hoyer,ICMLnonlingam}. It should be emphasized that such kind of reasoning
cannot be justified by referring to Occam's Razor only, i.e.,  the  principle to prefer simple models
if possible. The point that deserves our attention is to justify 
that Occam's Razor should  be applied to  {\it causal} conditionals
$P({\tt effect} |{\tt cause})$ instead of non-causal conditionals like $P({\tt cause} |  {\tt effect})$. 
Studying these asymmetries for time-series highlights the relation to commonly accepted asymmetries between
past and  future.

\vspace{0.3cm}
Acknowledgements: This work has been inspired by discussions with Armen Allahverdyan   
in a meeting that was part of the VW-project
``Quantum  thermodynamics: energy and information flow at the nanoscale''.

%\bibliographystyle{unsrt}
%\bibliography{../../../literatur/literatur}

\end{document}